\documentstyle[eqsecnum,epsf,aps,preprint]{revtex}
\begin{document}
\draft
\pagestyle{plain}
\preprint{IMSC-98/02/07 ,
hep-ph/9802390}
\def\be{\begin{equation}}
\def\ee{\end{equation}}
\def\ba{\begin{array}}
\def\ea{\end{array}}
\def\bd{\begin{displaymath}}
\def\ed{\end{displaymath}}
\def\bea{\begin{eqnarray}}
\def\eea{\end{eqnarray}}
\def\bra{\langle}
\def\ket{\rangle}
\def\a{\alpha}
\def\b{\beta}
\def\g{\gamma}
\def\d{\delta}
\def\e{\epsilon}
\def\ve{\varepsilon}
\def\l{\lambda}
\def\m{\mu}
\def\n{\nu}
\def\G{\Gamma}
\def\D{\Delta}
\def\L{\Lambda}
\def\s{\sigma}
\def\p{\pi}
\title{
LSND, solar and atmospheric neutrino oscillation experiments and
$R$-parity violating Supersymmetry
}
\author{Rathin Adhikari\thanks{
rathin@imsc.ernet.in}}
\address{The Institute of Mathematical Sciences,
C.I.T Campus, Taramani
Chennai-600 113, India
}
\author{Gordana Omanovi\`{c}\thanks{
gordana@ictp.trieste.it}} 
\address{
International Centre for Theoretical Physics
Strada Costiera 11, I-34013 Trieste, Italy
}
\maketitle
\date{14 October, 1998}

\begin{abstract}

With only three flavors it is possible to account for various
neutrino oscillation experiments.
The   masses  and mixing angles  
for three neutrinos can be determined  from   the available experimental data 
on 
neutrino oscillation and from the astrophysical arguments.     
We have shown here that such masses and mixing angles 
which   can explain   atmospheric neutrino anomaly, LSND result   
and   the solar neutrino 
experimental data, can be reconciled  
with the $R$-parity violating Supersymmetric Models through lepton number 
violating   
interactions.
We have estimated the order of  magnitude for some  lepton number violating
couplings. Our analysis  indicates   that the 
lepton    number violation is likely to be  observed   in              
near future    experiments.
From the  data on neutrino oscillation and  the electric dipole  
moment of electron,   under   
   some   circumstances  it is  possible to  obtain constraint   on 
the  
complex   phase  of  
some supersymmetry breaking  parameters  
in   $R$-parity  violating Supersymmetric  models.  

\end{abstract}

\section{INTRODUCTION}

Although in the Standard Model of electroweak and strong interactions
the neutrinos are massless  to all orders in perturbation theory, in
its extension,  the neutrinos  may acquire small   masses  with
see-saw type mechanism     in presence of sterile neutrinos.  Also
such masses   can  be present in  the minimal supersymmetric model
with  the    renormalizable lepton number violating  terms       in
the lagrangian. On the other hand, the astrophysical and cosmological
considerations also strongly  suggest the existence  of massive
neutrinos.  Presently, there are some possible evidences  \cite{Smi}  of massive
neutrinos   and the mixing  of  different  flavor of neutrinos
particularly  coming from  the anomalies observed in the solar neutrino flux  
   \cite{solar},  in the atmospheric neutrino production  \cite{{Fukuda},{atmos}}  
and in the  
 neutrino beams   from accelerators and reactors \cite{react}.
Although some of the evidences 
 like those coming from  solar neutrinos and   accelerator data has 
been explained \cite{patibabu} considering one massive and 
two nearly massless neutrinos  but  it is  in general
difficult to
fit various   neutrino data  considering three neutrinos as
particularly the first three evidences  are  best fitted by  three
different mass gaps  for neutrinos.
However, the conventional 
  approach to analyse various observed neutrino anomalies  in the experiments  
is to parametrise  
 those in terms  of   oscillation of  two neutrino states only. This assumption may 
     not  hold good while fitting several observed anomalies   simultaneously and 
consistent three flavor mixing scheme \cite{arc} for three neutrinos    to analyse
various data is essential. 
  Several authors \cite{{card},{scheck},{thun}}
have tried to fit   various experimental data on neutrino    oscillation 
  in the three flavor mixing    scheme.     It is     interesting to note  
   that  including 
the recent CHOOZ \cite{chooz}  and the SuperKamiokande \cite{Fukuda}  
   result  on
  neutrino oscillation  alongwith other  experiments in this direction,  
it  is  possible  to find the mass square differences  and the mixing angles  
for three neutrinos almost  
  uniquely  \cite{scheck,thun}.  Furthermore,  analysis  in three flavor mixing scheme
   indicates sizeable oscillations  
of electron neutrinos to tau neutrinos   that should be observed
  by the long baseline  neutrino experiments such as those utilizing a muon  
 storage ring at Fermilab 
 \cite{barger}.
  These analysis    \cite{{scheck},{thun}}   also indicate that the 
solar
neutrinos observed on earth should show no MSW effect        \cite{Smi}  as the   
 large  mass squared differences    has been considered  in those analysis.  
Precise measurement  of the multi-GeV, `overhead' ($\cos  \theta_z  \sim   1$)  
events  at SuperKamiokande  
will also be able to verify the three flavor mixing scheme   \cite{scheck,thun}
as the double ratio $R = {(N_\mu/N_e)}_{measured}/{(N_\mu/N_e)}_{no \;\;oscillation} $   
for electron and muon  for those events   
is somewhat less than 1  in the three flavor mixing scheme but this ratio     is 1 
   in the analysis with a single oscillation 
  process with small mass square differences for  neutrinos. 
However present SuperKamiokande data   are inconclusive  in this    low  
  $L/E$     region. 
Three   flavor mixing schemes  \cite{scheck,thun} give very good fit to the 
SuperKamiokande 
  data \cite{Fukuda}   for the double ratio 
for upward going events ( $\cos   \theta_z    < - 0.6   $   )  but  do not give very 
good fit to the  data on individual ratio for electron 
and  muon.  
However, the double ratios are less sensitive to systematic errors than the 
individual ratios. 
In these analysis  \cite{scheck,thun}   the LSND result  has been considered 
 as an   oscillation effect   rather than an unexplained background.  
In near future the BooNE  experiment   \cite{church}  will test the same channel
 of neutrino oscillation  as LSND with higher sensitivity and statistics.
Particularly the solutions for the mass square differences and the mixing angles 
  in the three flavor mixing scheme  as obtained in reference \cite{scheck}  
are not  
significantly contradicted by any  existing experimental result and the 
  conflicting evidences are below two sigma level. 
Future  various experiments on neutrino oscillation  and some of those 
 experiments with higher statistics 
 and lesser systematic  errors  will be able to verify the three flavor mixing 
 scheme  \cite{scheck,thun}  and   
 it will be  
       certain whether we really 
need  a fourth sterile neutrino \cite{bilenky}.  
At present,
we feel that three flavor mixing scheme   for neutrinos are very interesting 
 as   it has some specific predictions as mentioned  before   which
can be verified  by experiments and it tells about the mass squared differences and 
the mixing angles  almost uniquely.  

The uniqueness  of the mass square differences  and the mixing angles    
\cite{{scheck},{thun}}
for three neutrinos may have strong impact   on  physics beyond 
standard model in  the
way those constrain  the parameters of   other theories.  We like  to
study  such  impact on the  minimal $R$-parity  violating
Supersymmetric model  where neutrinos  can acquire mass. In
supersymmetric models, $R$-parity was introduced as a matter of
convenience  to prevent fast proton  decay.  It is now realised that
the proton lifetime can be made consistent  with experiment without
invoking discrete $R$-parity  symmetry.  If we do not impose conservation of  
$R$-parity in
the model, the  minimal  supersymmetric standard  model allows
the  following  $B$  and  $L$  violating terms  in the superpotenial
\footnote{ One may consider  another  term   $    \mu_{\alpha}  L^{\alpha}  
H_2   $  in the superpotential   \cite{nilles}.     However in general 
this lepton number violating term   can be rotated to the  first two 
terms in the  superpotential in (1.1) 
unless  a symmetry   of  $W$  does not commute  with the   
$SU(4)     $  
symmetry of   $  L^{\alpha}  $  rotations in the field space.
}
\begin{equation} 
W = \lambda_{ijk} L^i L^j {\left(E^k \right)}^c +
\lambda^{\prime}_{ijk} L^i Q^j {\left( D^k \right) }^c +
\lambda^{\prime \prime}_{ijk}{\left( U^i \right) }^c{\left( D^j
\right) }^c {\left( D^k \right) }^c      
\end{equation}
Here $L$ and $Q$ are the  lepton and quark  doublet
superfields, $E^c$ is the lepton  singlet  superfield,  and $U^c$
and $D^c$ are the quark singlet  superfields  and $i, j, k$ are the
generation indices.  In the above, the first  two  terms are  lepton
number violating  while the third term violates baryon number.  For
the stability   of  the  proton,  we  assume that only the $L$    -
violating first two terms in the superpotential is non-zero. One may
consider  some $Z_n  $ symmetry  to remove  $B$ -violating term in the
superpotential\footnote{   See   \cite{aulakh} for other alternative 
approaches   to forbid dimension
four as well as dimension five    $B  $  violating operators  but  
keeping        $  L  $  violating operators in the Lagrangian .}.  
As discussed later, $L$ -violating   couplings give rise to
masses for Majorana neutrinos   through  one loop diagrams as shown
in figure   1,       which  lead to neutrino oscillation phenomena.
In this work     we like to show that in the $R$  - parity violating Supersymmetric
 models, it is possible to obtain the  
required mass squared differences and the mixing angles for such massive 
 neutrinos to explain  LSND, solar
 and atmospheric neutrino oscillation experiments. 
   In our  analysis, it is possible to satisfy the  bound on the  effective 
mass for the Majorana neutrinos   obtained from the 
neutrinoless double beta decay experiment. We 
have estimated the magnitude    of    
 some
of the lepton number violating couplings $
\lambda_{ijk} $ and   $\lambda^{\prime}_{ijk}$   which is required to obtain
the appropriate  mass square differences and the mixing angles
for neutrino oscillation.
This kind of study was made earlier
\cite{masi}     in the  two flavor mixing scheme with the lesser available
neutrino data.  Very recently  some other studies \cite{pakvasa} also has
been made  to analyse  solar and atmospheric neutrino data  in the
context of Supersymmetric Models.
However in our work, unlike other works,
we have considered solar, atmospheric neutrino oscillation experiments  
as well as LSND data  to reconcile with   $R$ parity  violating
Supersymmetric Model. We have also discussed   the case for which neutrinos  
may  be considered as dark matter candidate.

There are stringent bounds on different $\l_{ijk}$ and
$\l^{\prime}_{ijk}$ \cite{drei} from low energy processes \cite{Barg}
and very recently the product of two of such couplings has been
constrained significantly from the   neutrinoless double beta decay
\cite{babu} and from rare leptonic decays  of the long-lived neutral
kaon, the muon and the tau as well as from the mixing of neutral $K$
and $B$ meson
\cite{Debchou}. In most cases it is found that the upper bound on
$\l^{\prime}_{ijk}$ and $\l_{ijk}$   varies   from $10^{-1}$  to
$10^{-2}$ for the sfermion mass of order 100 GeV. For higher sfermion
masses these values are even higher. In our analysis, it seems  that for sfermion 
 mass  of the order of 100 GeV  various $L$ violating  couplings  are less 
than $10^{-2}$.
Considering the values  of   $L$-violating  $\l^{\prime}$  couplings as obtained 
from 
our analysis and also considering the constraint obtained from the electric
dipole moment of  
electron  it is possible to obtain constraint on 
the  complex phase  
of some Supersymmetry parameters.   In  section II, we briefly 
discuss the constraints
on masses and mixing      for   three neutrinos  obtained from LSND,   solar and  
 atmospheric neutrino oscillation experiments.
 In    section III,
we discuss about masses of three   neutrinos in    $R$  violating 
    Supersymmetric   Model 
and show how it is possible to reconcile the masses 
 and the mixing angles  as obtained in the three flavor mixing scheme  
 with the  $R  $  violating  Supersymmetric Model.  We  present   the  required 
  values
of some  $L$ violating couplings which satisfy particularly various neutrino 
 oscillation experimental   data  and also satisfy         the  constraint 
 on the effective mass  for Majorana neutrinos  in the neutrinoless double beta  
 decay experiment.  
We compare these values 
with    the earlier constraint on such couplings. 
In section  IV,   we  discuss
  that under some circumstances     it   is possible   to get  
 constraint  
on the complex phase of some Supersymmetry parameters like $A$  -  parameter.  
In  section V, as concluding remarks we mention
the
possible implications of the obtained values of $L$ violating couplings 
in  
collider  physics  and cosmology.

\section{
    Constraint on neutrino    masses and  mixing   } 

  We  first mention here the necessary parameters for the three flavor
neutrino oscillation.  After that  following      references 
\cite{{scheck},{thun}}
we  
 shall  consider some specific values for the 
the masses and mixing as solutions  to satisfy  various available  
  experimental  data.   The
neutrino flavor eigenstate  are related to the mass eigenstate by

\be
\nu_{\alpha} =    \sum_{i} U_{\a i} \n_i
\ee
\noindent
where   $ U_{\a i}$   are the elements of a unitary mixing matrix  $
U    $ , $\n_{\a}     =     \n_{e, \m, \tau}   $ and      $ \n_i   =
\n_{1,2,3}  $.
\noindent
According to the standard parametrization  \cite{pdb}  of the unitary
matrix
\begin{eqnarray}
U_{\n}&=&\left(
\begin{array}{ccc} c_{12} c_{13}&s_{12} c_{13} & s_{13} \d
\\
-s_{12} c_{23} -c_{12} s_{23} s_{13} \d  & c_{12} c_{23}  - s_{12}
s_{23} s_{13} \d& s_{23} c_{13}
\\
s_{12} s_{23}  - c_{12} c_{23}    s_{13} \d    &        -c_{12} s_{23}
-     s_{12}  c_{23}       s_{13} \d     &   c_{23} c_{13}
\end{array}
\right).\nonumber\\
\end{eqnarray}
\noindent
where      $  \d = e^{i \d_{13} }    $  corresponds to the
CP-violating phase  (which will be neglected   here   )  and $c $ and
$s$ stand for sine and cosine of  the associated angle  placed  as
subscript.  The non -diagonal  neutrino mass matrix   $M_{\n}      $
in the flavor basis is diagonalised  by  the unitary matrix   $
U_{\n} $ as

\be
U_{\n}^T M_{\n} U_{\n} =  D_{\n}
\ee
\noindent
where  $D_{\n} $   is the diagonal mass matrix with the real
eigenvalues.  In the three generation neutrino mixing  scheme,  there
are two independent mass square differences.  These may be considered as
 $
\D_{21}   $ and   $
\D_{32}   $   where

\be
\D_{ij}    =   {\mid  m_i^2  - m_j^2 \mid }
\ee
\noindent 
and $m_i$ and $ m_j$ are the  neutrino  mass eigenvalues.   From the solar 
neutrino    deficit  one may consider     \cite{scheck}
\be
10^{-4} {\mbox eV}^2 \leq \D_{21}            \leq  10^{-3} {\mbox eV}^2
\ee
in which the lower limit is obtained from SuperKamiokande data 
\cite{Fukuda} and   
 the upper limit is obtained from the CHOOZ experiment \cite{chooz}.    
Keeping    in mind     both the atmospheric and LSND data,  
another mass square difference $\D_{32}$ can be considered as \cite{scheck}
\be
\D_{32}  \approx    0.3 {\mbox   eV}^2
\ee
in which the lower limit from the Bugey reactor constraint \cite{bugey} and the upper limit
 from the CDHSW   \cite{cdhsw} have also been considered. 
In the 
    one mass square difference  dominating the  other,   the three flavor
mixing scheme greatly simplifies  and   one can write  the     probability 
   of the  observation of  neutrino oscillation    in LSND and SuperKamiokande  in  
 terms   of    $\D_{32}$  and the     elements of  $U$    .   For the solar neutrino  
 experiment  the sine   squared terms containing   two  mass square differences 
in the expression    for  the  survival probability  of solar electron neutrino, 
can be averaged   for the flight length and the energy of the  neutrinos 
 observed on earth. In this case the probability can be written   in terms of  
 the elements    of $U$  only.  As the  probability of oscillation   in LSND 
  and  Superkamiokande  and the survival probability for solar electron neutrinos  
 are provided by the experiments,  one   can solve  
 for  the  three angles  by which    matrix $U$    in (2.2)  is defined.
The  results obtained by Barenboim    {\it et al}  \cite{scheck} show  that  
four set of solutions for three angles are possible. However, two set of  solutions
can be discarded  by considering   the   SuperKamikande  zenith angle  
 $(\cos \theta_z <   -0.6)$ behavior of  atmospheric neutrino    
 data for upward going events.  The other two allowed   
set of solutions for the three angles 
 as obtained in   reference   \cite{scheck}    are  
\be
\theta_{12}=54.5       ;  \qquad \theta_{13}   = 13.1;  \qquad \theta_{23}= 27.3 
\ee
\be
\theta_{12}=35.5;  \qquad \theta_{13}=13.1;  \qquad \theta_{23}= 27.3
\ee
   The  result obtained by Thun  {\it et al}   \cite{thun}   to satisfy solar   , atmospheric
 and LSND data   matches almost with the  second set of  solutions for  three angles
as mentioned in (2.8). 
In our analysis we  shall consider  either (2.7)  or (2.8)  for  the three angles  
  which specify the unitary matrix $U$ in     (2.2). 

The neutrino oscillation experiments give us information about the   
 mass squared differences of three neutrinos  in the three flavor mixing scheme 
as discussed above.  However, to know the mass of   different neutrinos 
we have to consider some other experiment.
 The masses   are generated  for  Majorana
   neutrinos  in     $R$   violating  Minimal Supersymmetric  Model,
so we have to consider the constraint coming  from neutrinoless double  beta decay.
This gives us an estimate for the masses of neutrinos.   
 The contribution of  Majorana neutrinos to the amplitude
of the neutrinoless double  beta decay    \cite{baly}   is
\bea
\mid{\cal M }\mid  \; =\;  \mid\sum_{i = 1, 2, 3 }  {  U_{ei}}^2
m_{\nu_i}\mid  < m(0 \nu \beta \beta   ) =  0.68   \;{\mbox eV
}
\nonumber
\eea
\noindent
To satisfy this constraint
  and keeping in mind that there are  some
uncertainties  in the calculation of the nuclear  matrix elements
one may consider different masses of the Majorana  neutrinos    of the
order of eV  or less \cite{nucl}.  
Considering        (2.5) and (2.6) alongwith this constraint it is found that 
there are two interesting possibilities for the masses of neutrinos.
 In one case, all three neutrinos have  
  almost degenerate mass and we may consider    
\be
m_2 \approx   1   {\mbox{eV}} 
\ee
then the   masses       for other two neutrinos are
\be
m_1 \approx 1 {\mbox eV};  \qquad     m_3  \approx   1.14    {\mbox     eV}
\ee
In another case, the masses of two neutrinos are nearly degenerate whereas the third 
 one is heavier and
 we may consider  
\be
m_2 \approx   3 \times  10^{-2}   {\mbox{eV}} 
\ee
then the   masses       for other two neutrinos are
\be
m_1 \approx 2 \times 10^{-2} {\mbox eV};  \qquad     m_3  \approx   0.55 
   {\mbox     eV} 
\ee

One may consider   the neutrinos as candidate
for the dark matter solutions also.  In that case, if  one
assumes   $\Omega  =1  $     and the  
energy density  of the neutrinos     $  \rho_{\n}  =  0.2  \rho_c   $ 
  where    $\rho_c$  is the critical density   in the  Big-Bang Model
    \cite{primack},  it is desirable to have the sum  of  the
neutrino masses   around  5   eV    and  one may consider  the nearly degenerate   
 three masses  of neutrinos  
given by (2.9) and (2.10). 

In our analysis  we shall consider  the above mentioned four interesting
possible solutions for masses and mixing angles - one set of solutions
from 
(2.7), (2.9) and (2.10), 
one        set of solutions from (2.8),
(2.9)  and (2.10), one set from  (2.7), (2.11) and  (2.12)
and the other set from (2.8), (2.11) and (2.12).    
We shall discuss  in the next section how all these solutions 
can be reconciled with   $R$ parity violating Supersymmetric Model. 

\section{
Neutrino mass   matrix in   $R$-violating
supersymmetric model and constraint on $L$ violating couplings}

  The trilinear   lepton number violating renormalizable term  in the
superpotential in    (1.1) generates Majorana neutrino masses
\cite{tomma,nilles}   through the generic  one loop diagram as
shown in figure    1 in which $s $   and ${\tilde s}   $  stand
for either  lepton and slepton or quark and squark respectively.  The
helicity flip  on the internal  fermion line is necessary and that
requires the  mixing  of    ${\tilde s}  $  and        $ {\tilde
s}^c$.      The contribution to the mass insertion as shown in figure
1, is proportional to  the  mass    $m_k $ of the fermionic
superpartner $s_k   $     of   $ { \tilde   s_k    }$   and  is also
proportional  to ${ \tilde   m   }$ (          $\sim  A $  $\sim  \m
$, where  $A$ and  $\m $  are  the  susy- breaking mass parameter)
.  The single diagram  in figure 1  that contributes to the Majorana neutrino
mass matrix    $m_{\n_i \n_j}      $    is

\be
m_{\n_i \n_j}        \approx      {\l_{jkn} \;   \l_{ink} \;  m_n
m_k { \tilde m }       \over   16 \pi^2 {\tilde {m_k} }^2   }
\ee

\noindent
when one considers  the lepton       and slepton  for  $s$ and
${\tilde s}  $     in the diagram in figure 1.  
 Both the diagrams in    (a) and (b) are to
be considered together and summed to evaluate  the neutrino mass
matrix element.   However for  $ i = j   $  and  $ k = n   $,    the
two diagrams coincide  and for that  only one is to be considered.
For quark and squark  in the diagram  the similar contribution will
be obtained. However, in that case,  the above contribution    is to be multiplied
by a color factor  3     and   $      {  \tilde   m_k  }$        in
the above equation  is to be considered as the  squark mass instead
of slepton mass and the $\l$ couplings  in (3.1) is to be replaced
by $\l^{\prime}$ couplings.

In constructing neutrino mass matrix we shall consider   the following
things.  Firstly       
we shall relate squark and slepton mass  as

\be
{\tilde  m_{slepton}   }^2             = {\tilde  m_{squark} 
}^2/K
\ee

\noindent
where  $  K  $  is a number  depending  on the various choices of
Supersymmetry parameters.   Different squarks  have almost degenerate mass and 
different  
sleptons also have almost degenerate mass    as    otherwise  there is 
  severe constraint from the flavor changing neutral current.  
There are        9 $\l  $ couplings and
27 $ \l^{\prime}  $      couplings  entering the neutrino mass matrix.
However, in writing each element of the neutrino mass matrix
 we shall  
consider only the leading term in terms of the magnitude of mass obtained from (3.1).
 We
shall consider the diagram with lepton and slepton in figure 1 and shall also
consider the diagram with squark and quark in figure 1 in each element
of the neutrino mass matrix for which two different types of $L$ violating  couplings
appear in each element.
 Under this consideration
only the following $L$ violating couplings appear in the neutrino mass matrix.
These are      $
\l^{\prime  }_{133  },
\l^{\prime  }_{233  } ,   \l^{\prime  }_{333  } ,   \l_{133  } ,
\l_{233  }, \l_{232},    \l_{132}$  couplings.
 The
notations for the first five couplings in later discussion will be     $
\l^q_e$  ,     $\l_{\mu}^q$, $\l_{\tau}^q$      ,  $\l_e^l$  ,
$\l_{\mu}^l$  
   respectively.  We are ignoring the effect of other couplings  in our analysis
and we are assuming   that the     $\l$ couplings   are not much hierarchical
  among themselves and  $\l^{\prime}$ couplings are also not much hierarchical
among themselves. 
At the end of      this section  we shall  make  a few qualitative comments  
   about considering  other    $   L     $   violating  couplings  in the mass matrix.  
We write the neutrino mass matrix as

\begin{eqnarray}
N&=&   a   \left(
\begin{array}{ccc}  K    m_{\tau}^2   {\l_e^l}^2   +  3 m_b^2 {\l_e^q}^2&
2 K    m_{\tau}^2   {\l_e^l}  {  \l_{\mu}^l   }   +  6 m_b^2 {\l_e^q}
\l_{\mu}^q   &         -2 K m_{\mu} m_{\tau} \l_{\mu}^l \l_{132}   +  
6  m_b^2    \l_e^q   \l_{\tau}^q\\ 2 K
m_{\tau}^2   {\l_e^l}  {  \l_{\mu}^l   }   +  6 m_b^2 {\l_e^q}
\l_{\mu}^q   & K    m_{\tau}^2   {\l_{\mu}^l}^2   +  3 m_b^2
{\l_{\mu}^q}^2 & - 2 K m_{\tau} m_{\mu}  \l_{\mu}^l \l_{232} +  
6  m_b^2    \l_{\mu}^q   \l_{\tau}^q
\\
-2 K m_{\mu} m_{\tau} \l_{\mu}^l \l_{132} + 
6  m_b^2    \l_e^q   \l_{\tau}^q    & -2 K m_{\tau} m_{\mu}  \l_{\mu}^l 
\l_{232} + 6  m_b^2    \l_{\mu}^q
\l_{\tau}^q   &   K    m_{\mu}^2   \l_{232}^2   +  3 m_b^2 {\l_{\tau}^q}^2
\end{array}
\right).\nonumber\\
\end{eqnarray}

\noindent
where
\be
a    =     {  {\tilde m  }    \over  16  \pi^2  {  \tilde  m_s }^2  }
\ee

\noindent
and $ {\tilde m_s   }      $     is the  almost   degenerate squark
mass.  The eigenvalues for  this matrix correspond to three masses  $m_1$,
$m_2$ and $m_3$ for three Majorana neutrinos.     
 We can write the diagonal
mass matrix as
\bea
{D_{\nu}}&=&\left( 
\begin{array}{ccc}m_1&0&0\\0&m_2&0\\0&0&m_3
\end{array}
\right)
\eea

\noindent
All the elements of this diagonal mass matrix can be written by considering 
particular set of solutions for the masses 
from the  earlier section. 
The  unitary matrix $U_{\nu}$ 
in (2.2) diagonalising     the non-diagonal neutrino mass matrix in the  
  flavor basis is also known to us if we  consider particular set  of  solutions 
  for the three angles from the earlier section. As both  $D_{\nu}$ and  $U_{\nu}$ 
  are known  we can obtain the non-diagonal mass matrix   $M_{\nu}$      
 in the flavor basis using
 the relation
\be
U_{\nu} D_{\nu} U_{\nu}^T  =  M_{\nu}.   
\ee
\noindent
So for particular set of solutions for the masses and the 
 mixing angles discussed in the earlier section,  all the elements   of   
$M_{\nu}$  are  
 known.   However, this  $M_{\nu}$ is   equal to  $N$ 
which is also
the non-diagonal mass matrix expressed in terms of different $L$ violating 
 couplings.  So we write
\be
N  =   M_{\nu}
\ee
From  (3.7) we get six equations  for the $L$ violating couplings
 : 
\be
 K    m_{\tau}^2   {\l_e^l}^2   +  3 m_b^2 {\l_e^q}^2 =     M_{\nu}(1,1)/a
\ee
\be
2 K    m_{\tau}^2   {\l_e^l}  {  \l_{\mu}^l   }   +  6 m_b^2 {\l_e^q}
\l_{\mu}^q  =  M_{\nu}(1,2)/a
\ee
\be
- 2 K m_{\mu} m_{\tau}      \l_{\mu}^l   \l_{132} + 
       6  m_b^2    \l_e^q   \l_{\tau}^q  =  M_{\nu}(1,3)/a
\ee
\be
K     m_{\tau}^2   {\l_{\mu}^l}^2   +  3 m_b^2
{\l_{\mu}^q}^2 = M_{\nu}(2,2)/a 
\ee
\be
 -2 K m_{\tau} m_{\mu}  \l_{\mu}^l \l_{232}
+
6  m_b^2    \l_{\mu}^q   \l_{\tau}^q
= M_{\nu}(2,3)/a
\ee
\be
 K    m_{\mu}^2   \l_{232}^2   +  3 m_b^2 {\l_{\tau}^q}^2
= M_{\nu}(3,3)/a
\ee
after comparing the elements  (1,1), (1,2), (1,3), (2,2), (2,3) and (3,3) respectively. 
However, there are seven $L$  violating couplings involved in these six equations. 
So we shall consider the possible value of one of the $L$ violating couplings  
 from some other experiment instead of neutrino oscillation experiments for  solving 
  above six equations to find six $L$  violating  couplings.  
We shall consider  particularly   
some value  of   $\l_{232}$
lower than  ${0.006 {\tilde    m_s}   \over \sqrt{\tilde m}}$   
which is allowed     after considering the constraint from lepton universality 
\cite{{drei},{Barg}}.
From these six equations we  can determine the values of  
six $L$ violating couplings
for which   it is possible to reconcile  LSND, solar and atomspheric      neutrino  
oscillation experimental data with the $R$ parity violating Supersymmetric Model.   

To determine
$M_{\nu}$ we  first consider (2.7), (2.9) and (2.10) for the masses and the mixing
angles. From (2.9) and (2.10) we get a specific  
$D_{\nu}$  in (3.5), and from (2.7) we get
a specific $U_{\nu}$ in (2.2). Using relation (3.6) we obtain the following form of $M_{\nu}$ (
Here and in later discussions to obtain $M_{\nu}$ we shall consider $m_1$, $m_2$ and 
  $m_3$ in (3.5)  in eV unit) :
\bea
{M_{\nu}}&=&\left(
\begin{array}{ccc}1.00712&0.0143034&0.0274612\\0.0143034&1.02782&0.0542459\\
0.0274612&0.0542459&1.10499
\end{array}
\right)
\eea
We take
$m_b =  4.3   \times 10^9  $ eV ,  $m_{\tau}  =  1.777
\times 10^9  $  eV  and   $  m_{\mu}  = 0.105658  \times 10^9 $  eV
and solve (3.8)-(3.13) 
after considering a specific $M_{\nu}$ in (3.14).
For various allowed real values of $\l_{232}$  lower than that mentioned earlier,
the   solution for other six  $L$ violating
couplings   do not change by an order. We present below the the values of
these couplings considering  $ \l_{232}$  in the range
$(10^{-6}-10^{-3})   \; {{\tilde m_s} \over \sqrt{\tilde
m }}$
(Here and in
later discussions ${\tilde m_s}$ and ${\tilde m}  $ stand for the corresponding 
magnitude
in GeV unit):
\be
\l_e^q \approx  { \;5.3     \times  10^{-5}
{\tilde m_s   }      \over           {\sqrt{\tilde m }}  }
, \qquad
\l_{\mu}^q   \approx   { \;1.3    \times  10^{-6}
{\tilde m_s   }      \over           {\sqrt{\tilde m }}  }
,   \qquad
\l_{\tau}^q  \approx    { \;5.6   \times 10^{-5}
{\tilde m_s   }      \over           {\sqrt{\tilde m }}  }
\ee
\be
\l_e^l  \approx  {7.3   \times  10^{-6}
{\tilde m_s   }      \over           {\sqrt{     K    {\tilde m }}
}}
, \qquad \l_{\mu}^l      \approx          {2.3    \times   10^{-4  }
{\tilde m_s   }      \over           {\sqrt{     K    {\tilde m }}}
}, \qquad
\l_{132}      \approx          {4.0    \times   10^{-3  }
{\tilde m_s   }      \over           {\sqrt{     K    {\tilde m }}}
}
\ee
\noindent
Another set of real  solutions for various $L$ violating couplings  for the above   
  case is given below :
\be
\l_e^q \approx  { \;5.3     \times  10^{-5}
{\tilde m_s   }      \over           {\sqrt{\tilde m }}  }
, \qquad
\l_{\mu}^q   \approx   { \;1.3    \times  10^{-6}
{\tilde m_s   }      \over           {\sqrt{\tilde m }}  }
,   \qquad
\l_{\tau}^q  \approx    { \;5.6   \times 10^{-5}
{\tilde m_s   }      \over           {\sqrt{\tilde m }}  }
\ee
\be
\l_e^l  \approx  {4.2   \times  10^{-6}
{\tilde m_s   }      \over           {\sqrt{     K    {\tilde m }}
}}
, \qquad \l_{\mu}^l      \approx          {2.3    \times   10^{-4  }
{\tilde m_s   }      \over           {\sqrt{     K    {\tilde m }}}
}, \qquad
\l_{132}      \approx          {3.9    \times   10^{-3  }
{\tilde m_s   }      \over           {\sqrt{     K    {\tilde m }}}
}
\ee
\noindent
We have ignored overall + or - sign for the solutions (here and in later
cases also) for different   $L$ violating 
couplings.  Although there are different set of solutions possible but
if we ignore the small changes in the higher decimal places  for  different solutions
then   mainly the   
two    set of solutions   are found to differ to some extent
from each other particularly for  the  value of    $\l_e^l$  
and $\l_{132}$ and those two sets of solutions are presented 
above. 

It is important to note here that there is almost no change    of the 
$\l_{132}$ value for various real $\l_{232}$ value in the range 
$(0.0   - 10^{-3})    
{{\tilde m_s   }      \over           {\sqrt{     K    {\tilde m }}}
}$  and the
 two values of $\l_{132}$  in (3.16) and (3.18) 
are very near to the  upper bound obtained
from the   experimental value  of  $ R_{\tau}  =   { \G (\tau \rightarrow
e \nu {\bar \nu}) \over \G (\tau \rightarrow \mu \nu {\bar \nu})}$ 
\cite{drei,Barg,aleph}. This indicates that  there is possibility to see the 
lepton universality
violation in future experiments. Same comment is also true for $\l_{232}$ coupling as
almost same real solutions  for various $L$ violating couplings  exist for the 
 higher allowed   
value of  $\l_{232}$ coupling also. Of course the statement is based on the present 
 neutrino oscillation data and considering neutrino as Majorana particle 
, the main contribution  in
 the neutrino mass matrix coming  from   the earlier mentioned seven  $L$
  violating couplings and the $L$  violating couplings considered  here being
 real.  
Although in obtaining    the  solutions for     $   L           $  violating 
couplings we have considered     here almost degenerate  mass for  three neutrinos ,
    but such    higher values of   $\l_{132}$ or     $\l_{232} $    are   possible for  
    hierarchical nature  of the masses of neutrinos    also,   as can be seen      
in the later part of    our analysis. 
If we consider complex or imaginary value of $\l_{232}$ coupling considering 
the experimental upper bound mentioned earlier    it is possible to obtain 
complex solutions for other six $L$ 
violating couplings   from (3.8)-(3.13) for   $M_{\nu}$ in (3.14). For brevity, we
are not presenting   those solutions of  various  $L$ violating couplings
for this case of masses and mixing angles and for other cases also.
 However, at the end of this section we shall make a few general  
   remarks  on the  complex solutions for  these $L$ violating couplings.
Existence of the possible solutions for $L$ violating couplings  in
(3.15) and (3.16) or (3.17) and (3.18) indicates that considering  the almost 
degenerate 
mass neutrinos
(which may be candidate for dark matter also)  as mentioned   
in (2.9) and (2.10), it is possible
to reconcile LSND, solar and atmospheric neutrino oscillation  data with the    $R$
 parity violating Supersymmetric  Model.  

Next, we consider the other possible
solutions for the mixing angles as stated in (2.8) and consider again the
almost degenerate mass of neutrinos  as mentioned in (2.9) and (2.10). In this case, 
we get the following form of $M_{\nu}$ after using (3.6) :
\bea 
{M_{\nu}}&=&\left( 
\begin{array}{ccc}1.00704&0.0143116&0.0274772\\0.0143116&1.02788&0.054211\\
0.0274772&0.054211&1.105
\end{array}
\right)   
\eea  
Like earlier case, we again solve equations (3.8)-(3.13) for specific $M_{\nu}$  in
 (3.19)
and obtain the solutions for six $L$ violating couplings. All the solutions
in this case are  approximately same  as (3.15)-(3.18). So the different       
set of choices for the mixing angles in (2.8) do  not lead to significant change    
 in the values   of $L$  violating   couplings. 

We shall consider next the hierarchical neutrino masses as mentioned in (2.11) 
and (2.12).
 For the three mixing angles we consider (2.7). As before using (3.6) we  obtain the
following form of $M_{\nu}$  :  
\bea 
{M_{\nu}}&=&\left( 
\begin{array}{ccc}0.0534391&0.0569347&0.10027\\0.0569347&0.127333&0.202493\\
0.10027&0.202493&0.417771
\end{array}
\right)   
\eea  
Solving (3.8)-(3.13) for specific $M_{\nu}$  in   (3.20) we obtain the following 
real solutions for six $L$ violating couplings.  We present below the values of
these couplings considering  $ \l_{232}$  in the range  $(10^{-6}-10^{-3} )  \; 
{{\tilde m_s} \over 
\sqrt{\tilde
m }}$.
\be
\l_e^q \approx  {8.5     \times  10^{-6}
{\tilde m_s   }      \over           {\sqrt{\tilde m }}  }
, \qquad 
\l_{\mu}^q   \approx   {8.3    \times  10^{-6}
{\tilde m_s   }      \over           {\sqrt{\tilde m }}  }
,   \qquad
\l_{\tau}^q  \approx    {3.4  \times 10^{-5}
{\tilde m_s   }      \over           {\sqrt{\tilde m }}  }
\ee
\be
\l_e^l  \approx  {3.7   \times  10^{-5}
{\tilde m_s   }      \over           {\sqrt{     K    {\tilde m }}
}}
, \qquad \l_{\mu}^l      \approx          {7.2    \times   10^{-5  }
{\tilde m_s   }      \over           {\sqrt{     K    {\tilde m }}}
}, \qquad
\l_{132}      \approx          {1.8   \times   10^{-3  }
{\tilde m_s   }      \over           {\sqrt{     K    {\tilde m }}}
}
\ee
Another set of real solutions for various  $L$ violating couplings for the above 
case  is given below :
\be
\l_e^q \approx  {1.2     \times  10^{-5}
{\tilde m_s   }      \over           {\sqrt{\tilde m }}  }
, \qquad
\l_{\mu}^q   \approx   {8.3    \times  10^{-6}
{\tilde m_s   }      \over           {\sqrt{\tilde m }}  }
,   \qquad
\l_{\tau}^q  \approx    {3.4  \times 10^{-5}
{\tilde m_s   }      \over           {\sqrt{\tilde m }}  }
\ee
\be
\l_e^l  \approx  {5.2   \times  10^{-6}
{\tilde m_s   }      \over           {\sqrt{     K    {\tilde m }}
}}
, \qquad \l_{\mu}^l      \approx          {7.2    \times   10^{-5  }
{\tilde m_s   }      \over           {\sqrt{     K    {\tilde m }}}
}, \qquad
\l_{132}      \approx          {1.2   \times   10^{-3  }
{\tilde m_s   }      \over           {\sqrt{     K    {\tilde m }}}
}
\ee
\noindent
So   it is seen that considering hierarchical nature of the masses of  
neutrinos   also it is possible to reconcile  LSND,    solar and atmospheric neutrino  
  oscillation experiments  with the $R$  violating Supersymmetric Model.  In this
  case    only  thing to note here    is that   $\l_{132}$  is slightly lower than 
 the earlier cases    however not far from the present   experimental bound  obtained 
from 
 the    lepton universality violation \cite{drei,Barg,aleph}.       

Next we shall consider the hierarchical mass pattern of neutrinos like
earlier case but consider the mixing angles as presented in (2.8). In that  case,  
using (3.6) we get the following form of   $M_{\nu}$ :  
\bea 
{M_{\nu}}&=&\left( 
\begin{array}{ccc}0.0503506&0.0572644&0.100908\\0.0572644&0.129869&0.201098\\
0.100908&0.201098&0.418324
\end{array}
\right)   
\eea  
Like before we solve (3.8)-(3.13) for $M_{\nu}$ in (3.25) and consider  the 
same range  for $\l_{232}$ like earlier cases. 
In this case,  the solutions for $L$ violating couplings  are almost same as
before with  hierarchical masses of neutrinos  and we are not presenting those 
 solutions seperately. 

If the future neutrino oscillation experiments with higher sensitivity  and 
  more data support the  three flavor mixing scheme as mentioned   in  
  section   II and the  $L$ violating couplings are real, it is expected that 
experiments on  
lepton universality violation 
 in future will   find signal for the values of $\l_{132}$ or $\l_{232}$ 
couplings at the level required by our analysis. However, 
 if no signals are found   for those values of $L$ 
 violating ocuplings-  particularly  $\l_{132}$ 
 coupling   then the explaination for
that may be the following.   In that case  normally it will be expected
that $\l_{132}$ coupling is very small.   However then if one considers
again  those six equations (3.8)-(3.13)  considering  $\l_{132}$ as 
effectively zero it is found that for the various
cases  for masses and mixing angles,   real values  of   
  $\l_{232}$  has to be  always of the
order of    ${    10^{-3  }
{\tilde m_s   }      \over           {\sqrt{     K    {\tilde m }}}
}$.  However if the signal for $\l_{232}$  is also not seen through
 $\tau$-universality   violation,    it will be necessary to   check the role of other 
couplings for the 
 analysis  of  neutrino masses and mixing angles.     As under this    
circumstance,  $\l_{132}$ and $\l_{232}$  will be smaller,     we may consider  the 
terms next to the  leading order in mass    in the various elements
 of the neutrino 
mass matrix in (3.3). 
As the mass factor associated with those other non-leading contribution 
will be less  in magnitude, it is expected that  the magnitude of some other 
coupling should be 
somewhat higher   like $\l_{132}$ 
to reproduce  the similar forms of  $M_{\nu}$ mentioned
earlier  and the lepton number violation, in that case, should be observed through
that coupling.
Depending on the  results of future experiments  on neutrino oscillation,
tau universality violation etc., the analysis with
 other  such couplings may be    important.

We like to make a few remarks on the complex  
solutions for various $L$  violating   
couplings. 
Earlier in obtaining all the  above-mentioned solutions for different $L$ - violating
couplings  
we have considered value of $\l_{232}$  coupling in the range $(10^{-6}-10^{-3})  
{ \tilde m_s 
\over   \sqrt{K {\tilde m}}}  $.    However,  if one considers   the  value of
  $\l_{232}$ in the range $(4.0-6.0) 10^{-3}  { \tilde m_s  
\over   \sqrt{K {\tilde m}}}  $    which is very near to the experimental upper 
bound   \cite{drei},    
then from the
equations  (3.8)-(3.13) considering different form of   $M_{\nu}  $ as 
 mentioned earlier,  one will obtain the  complex solutions for various
$L$ violating couplings.
Furthermore,   if one considers imaginary or complex value  of   $\l_{232}$ in that
 case also one will obtain  the  complex solutions for various $L$ violating couplings.
Depending on the various choices of  the values of   $\l_{232}$   one may obtain   
from (3.8)-(3.16)  the various solutions for different $L$  violating couplings
with various possible complex phases. For brevity, we have not 
  presented various possible complex solutions. For some complex phases
  the solutions may not be allowed depending on constraint
particularly  from  the  value of the electric dipole moment of electron.  We
have discussed this in the next section.    

The solutions for all   $ \l $ and  $ \l^{\prime}  $ couplings are
obtained in terms of the  parameters    $       {\tilde m }    $, $
{\tilde m_s }    $   and   $ K $.  Here $       {\tilde m }    $
is the Supersymmetry  breaking mass parameter  which is expected to lie in the
range    of    O(100 GeV )     to   O(TeV ).  
To
compare our constraint  on   $ \l $ and   $ \l^{\prime}  $ couplings
with  the earlier constraints     \cite{drei}   we shall consider
$K  \approx 1 $ which means slepton mass does not differ much  from the squark
mass  and also  shall consider  
both squark mass and   $  {\tilde m }  $  of  O(100 GeV  ).
We shall  consider those solutions of $L$ violating
couplings for which complex phases are negligible.  
The earlier constraint from the tau universality
violation  \cite{Barg} is   $\l_{132}  \leq 0.06$,     $\l_{232} \leq 0.06$ and  
${\l_{233}}     =       \l_{\mu}^l    \leq
0.06$  which in our analysis from neutino oscillation,    are found  to  be
O(0.01-0.04),  O(0.01) \footnote{When we consider         very small value of 
 $\l_{132}$ which 
 can be neglected  in the  neutrino mass matrix in (3.3),  in that case 
  we get this solution for $\l_{232}$ from   the equations (3.8)-(3.13). Otherwise 
various     solutions for   $\l_{232}$ are possible as mentioned earlier.} and  
O( 0.0007-0.0023) respectively.  The earlier constraint  from    $ R_l  =
\G_{hadron}(Z^0)  /
\G_l(Z^0)  $    \cite{ellis}   
$   \l_{333}^{\prime } =  l_{\tau}^q   \leq   0.26$ and   $
\l_{233}^{\prime}   =   l_{\mu}^q      \leq 0.39   $    which in our
analysis   are $O(0.0003  -0.0005)$  and   $O((1-8) \times 10^{-5})$ respectively. 
The earlier
constraint  on  $    \l_{133}^{\prime}    = \l_e^q   $  obtained from
the    constraint from  neutrino mass       \cite{hall}  is $  \l_e^q
\approx 0.002  $ and  $
\l_{133}       =    \l_e^l   \approx   0.004  $  which in our analysis  are
O(0.00008-0.00053) and    O(0.00004-0.00037)   respectively.  So our
analysis indicates somewhat lower values of  various $L$ violating 
 couplings  than the  upper bound on these couplings obtained from other experiments.  
Furthermore, if one considers    $     {\tilde m
}  $    to  be nearer to  TeV  region then    these values 
of $L$ violating couplings will be further
lowered.
The upper bounds on these couplings obtained
from the experimental data on neutral currents ,  $  \beta  $ decay
\cite{Barg},  muon decay $(\mu
\to  e   \g  ,         \mu  \to   { \bar e } e e   )$ \cite{Barn},
or tau decay $(\tau    \to  \mu       \g  ,  \tau  \to  e  \g   )
$  \cite{masi1} etc.  are somewhat higher than the values required in our analysis.  
\section{   Constraint  on complex phase  of Supersymmetry breaking parameter}

In the  Standard Model  the  electric dipole moment  of  electron is 
much smaller  than   their present experimental bound   $d_e   <  10^{-26} 
e$cm  \cite{abdul}.    So the new  sources of  CP violation  which  
occurs  in   the   supersymmetric model    can be studied  on the  
basis  of  electric dipole moment of    electron  \cite{kiz,aoki,frank}.  
In the  minimal  supersymmetric standard model  apart from the Yukawa  couplings 
there  are several 
complex parameters  like three gaugino masses  corresponding to  SU(3), SU(2) and
U(1) groups, the mass parameter 
$m_H$ in the bilinear term in the  Higgs superfields in the superpotential 
, dimensionless parameters   $A$ and   $B$  in the trilinear and 
the bilinear terms of  
the scalar fields.            With   suitable redifinition  of the  fields, some  
   of these                 parameters can be made real  but  in that case 
some others can not be  
  made real like     $A$ parameter   \cite{aoki}.     The  complex  $A$
 will contribute  to  the electric dipole moment (edm) of electron.    
Furthermore,   if  
we  consider 
 the  complex    $   \l^{\prime}$  couplings,   the complex phase  associated 
with those will  also     
contribute to  edm of 
 electron.
   There will be   various diagrams  in the    $R$-parity   violating Supersymmetry
  for  the
 edm of electron  \cite{frank}.   But the significant contribution to edm  comes 
 from the one loop  diagram  containing top quark in the  loop as  shown in Figure- 2.  
  There will be diagram  containing massive neutrinos in the loop  .  However,  the 
 masses  of neutrinos  are quite small in our discussion and we are ignoring 
those types
 of diagrams  for our discussion as there will be lesser contribution to the 
 edm of electron.    In terms of complex phases we can write  
$A_f$ and    $\l^{\prime}_{ijk}$ as  
\be
  A_{u,d}  =   \mid A \mid    \;  exp \; (i \a_{A_{u,d}}   )  ,   \qquad    
   \l^{\prime}_{ijk}  =   \mid \l^{\prime}_{ijk}  \mid \;  exp \;  (i \b).  
\ee
  and  the  mixing angle for  the left and the right squark  in the familiar way as
\be
  \tan 2 \theta  =  2       \mid A_u  \mid \; m_u/(\mu_L^2 - \mu_R^2  ).
\ee

\noindent
Following reference  \cite{frank} and assuming 
 different   $\l^{\prime}_{ijk}$ 
       containing  another complex phase as  mentioned in  (4.1)   
we can write the edm of electron from figure 2  as

\bea
 d_e  \approx   &-&    \sin \theta \;\cos \theta 
\left( ( \cos^2 \b -\sin^2 \b) \;\sin \a_{A_d} \;+  \cos \b      
\;\sin \b  \;\cos \a_{A_d}    \right) 
\nonumber \\  &\times& \;{\mid\l^{\prime}_{1jk}\mid}^2  
\;\;    { 2 \;\mid A_{d}\mid    \over  3  {\tilde m_s}^3  }  \;m_{u_j}\;  
\left[ F_1 (x_k)  
   + 2 F_2 (x_k) \right]      \;10^{-17}    \;e \;{\mbox cm}
\eea

\noindent
where   $x_k  = {\left({  m_{d_k}     /  {\tilde m_s} }\right)}^2$  
and  the loop integrals  $F_1$  and  $F_2$ are expressed in terms of   $x_k$ as   

\bea
F_1(x_k)  = { 1 \over  2 {(1-x_k)}^2}  \left( 1+ x_k + { 2 x_k \;\ln x_k   
\over 1 - x_k} 
    \right) ,  \nonumber      \\
F_2(x_k) =   { 1 \over  2 {(1-x_k)}^2}  \left( 3-   x_k + { 2  \;\ln x_k   
\over 1 - x_k}
    \right)  .  
\eea

\noindent
$A_d$  and  $      {\tilde m_s}$  in  (4.3) correspond to the 
  magnitude of  those quantities  expressed in GeV.  We are particularly 
interested for $j=3$ and  $k=3$  case  in (4.3).      We got a   solution 
 for   $\l_e^q  =  \l^{\prime}_{133}  $ in section  III to explain the neutrino physics  
 data.    So we like to constrain  here particularly the complex phases  associated 
with  $A$ in  (4.3).     Considering 
    $  \l_e^q$  as real or complex and writing it as
$ C  {\tilde m_s}  /  {\sqrt  {  
\tilde m}}
$  in the form obtained in section  III   (where $C$ is some value depending
on the  type of solutions),    $A$ and 
 $   {\tilde  m_s }     $     both   from  100 GeV  to  1 TeV 
it is  
found that   the  constraint on the complex phase $\a_{A_d}$  and   $\b$  is  
\bea
\left( ( \cos^2 \b -\sin^2 \b) \;\sin \a_{A_d} \;+  \cos \b
\;\sin \b  \;\cos \a_{A_d}    \right)  
\sin  \theta   \;  \cos \theta 
\nonumber  \\  
\stackrel{\displaystyle <}{\sim}  {(2   - 16)  \times 10^{-14} 
 \over   {\mid C  \mid}^2} 
\eea

\noindent
In the case for which there will be  no contribution to  edm  in     (4.3)  
     is     $  \b  =   -   \a_A  /2    $. For the complex
solutions  of     $\l_e^q$    for which      $\mid   C   \mid  $   is not less 
  than    about $   10^{-6}  $ we can make the following statements.
For  $\b  \approx \pi/4  $     
and    $  \a_{A_d} \approx \pi/2   $  the 
 above  inequality can  be satisfied   for  any value  of    $\theta$.   
However such   a large phase for     $A$  is  
   not possible  as    the edm of electron will  get contribution from other diagrams
   involving  charginos and neutralinos   at the  one loop level \cite{kiz} cancelling 
  this possibility.  So 
   $\b  \approx \pi/4  $   is  not  possible  for any value  of    $\theta$.  
So those    set of complex solutions for   different   $L$  violating 
couplings  should
 not be considered  when the complex phase associated with   $\l_q^e$ is found to 
 be approximately   $\pi/4$ and $\mid C   \mid  $     satisfies the  above condition.
  Let us consider that     $ \l^{\prime}_{1jk}     $  are real   and      $  \theta = 
\pi/4$    in that case 
  it is seen from section III  that    $ C \approx 10^{-5}$ and   
the complex  phase  for  the   $A$ parameter   
        $   \a_{A_d}   <     3.2 \times 10^{-3}$.
 Without any   specific   choice 
  of  the  mixing angle  $\theta$      one   can constrain  only the combination 
  of    $\b$,     $\a_A$ and   $\theta$   as shown in (4.5).  

\section{
 Conclusion}

We have shown here  that in the minimal supersymmetric model with
$R$-parity violating trilinear term in the superpotential in (1.1)
it is possible to obtain the appropriate
mass square differences and the mixing angles   as required to  
  explain the LSND,  atmospheric and solar neutrino oscillation experimental data
in the three flavor mixing scheme for neutrinos. 
The validity of the three flavor mixing scheme  can be verified in the   near  
 future experiments on neutrino oscillation  as mentioned in the introduction.  
The masses for three Majorana neutrinos are generated at the one loop level 
 as shown in Figure-1  and it is possible to satisfy the  constraint on the
masses and the mixing angles  coming from   the neutrinoless double beta decay.
In each element of   the neutrino mass matrix in (3.3)   we have considered 
two leading terms  in terms of the magnitude of masses in (3.1)  
 coming from the diagram with
slepton and lepton and also coming from the diagram with quark and squark in  Figure-1.
Under this consideration it is interesting to note that for real
values of various $L$ violating couplings  at least one of the couplings either 
$\l_{132}$
or   $\l_{232}$ is expected to be quite high and very near to the experimental  
upper bound  coming from the     $\tau$             -universality violation 
\cite{drei,Barg,aleph}. Apart from these two particular couplings for some of the 
$L$  
violating couplings the magnitude are such that
it might be possible  to observe such $L$ violating interaction  
 at the Tevatron  or at  HERA.
 At the Tevatron after squark pair  
  production those squarks will decay  to  LSP (say neutralino) and which 
  will decay   via
$   L^i L^j {E^k}^c    $
operator   giving  multilepton signal   \cite{DP}.  At HERA  one can see  
  $R$  violating   Supersymmetry  signal  for
    $L^i Q^j {D^k}^c  $        operator
\cite{drei2}  through resonant squark production   and its subsequent decay  
  to electron or positron  and neutrino  giving the signal of  high    $p_T$
 electron or high   $P_T$  positron  or missing  $p_T$  for neutrino.
The  basic requirement for the observation of such signal  is  that LSP  has 
 to decay inside the detector   and this puts bound      \cite{DP,drei2} 
\bea
\l ,    \;    \l^{\prime}   \stackrel{\displaystyle >}{\sim}
   10^{-5}   \;    {\left(   { m_{{\tilde l},{\tilde
 q}}
   \over   100     \;{\mbox  GeV}}      \right)}^2
\nonumber
\eea
where $m_{{\tilde l},{\tilde
 q}}
$
   stand for the squark  and the slepton mass. From the above condition 
it is seen that if we consider   $  {\tilde m}$ of
  the order of  squark mass $  {\tilde m_s}$ in that case for squark mass
or slepton mass of the order of 300 GeV,      it may be possible to   observe 
   $L   $ violating signal for those couplings  discussed in section III  
for which   $\l, \;\l^{\prime}   > 5 \times   10^{-6}    {{\tilde  
  m_s} \over  \sqrt{\tilde m}}$; 
whereas for squark or slepton  mass of  the order of TeV   the condition  is
$  \l   ,  \;  \l^{\prime} >    3 \times 10^{-5}    {  {\tilde m_s}   \over  
\sqrt{ {\tilde m}}}$.
So for various  couplings
considered in our analysis in section III,   it  might be possible to 
observe $L$ violating   signal.    

If one considers the
baryogenesis in the early universe at the GUT scale, after the
generation  of asymmetry to satisfy the out of equilibrium condition
one requires  $L$ violating couplings $  \sim 10^{-7}     $  for
squark mass   from 100 GeV to 1 TeV  range
\cite{camp}   which is  significantly smaller than the values of some of the 
couplings 
 obtained in section III.
So if one likes to satisfy  the neutrino physics experimental  data  in the three flavor
mixing scheme, it 
       seems   in the  $R$
violating Supersymmetric scenario    the generation of the  baryonic  asymmetry
near the electroweak     scale  is more favored  where the constraint
on $L$ violating couplings   are not so severe     \cite{masi2}.
We have shown  that in the $R$ violating Supersymmetric   models      neutrino  
  can be considered as dark matter candidate also.   
Our  analysis  also indicates  that  to satisfy   various  
  experimental  data  on neutrino oscillation   
    ,  the lepton  number violating couplings    
 are constrained in such a way  that some combinations  of   left and right squark
    mixing angles and the complex phases of  some  
   Supersymmetry parameters - particularly that    of     $A$  parameter
are   
constrained.

\hspace*{\fill}

\noindent
{\large \bf       Acknowledgment  }

One of us (R.A) likes to thank E. Lisi for nice discussion on neutrino
 oscillation experiments. Both of   us  like to thank   G.  Senjanovic  and F. Vissani 
for discussion  at the  early 
 stage of this  work. One of  us (R.A)  wishes to thank
A. Smirnov also for
discussions  during the Extended Workshop
on the Highlights in Astroparticle Physics    held at ICTP during 15-th Oct.- 
15-th  Dec.,'97.

\newpage
 
\begin{figure}[htb]
\mbox{}
\vskip 3.0in\relax\noindent\hskip -0.17in\relax
\includegraphics{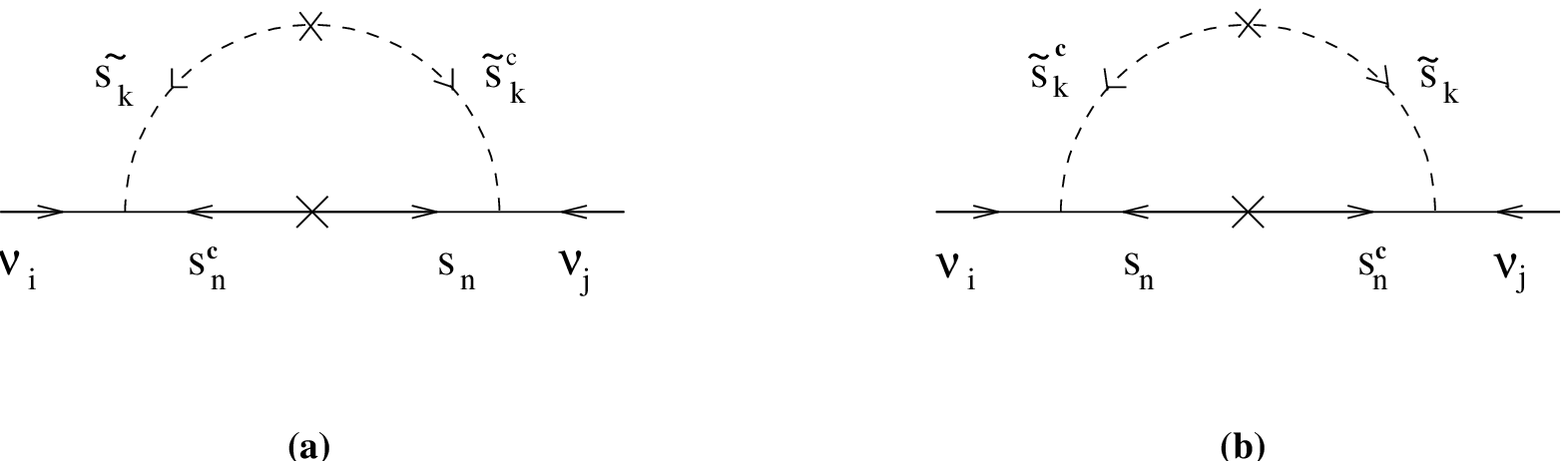} \vskip .25in
\caption{   One loop  diagram  involving $L$-violating  couplings  
  generating neutrino 
mass.}
\end{figure}
\newpage
 
\begin{figure}[htb]
\mbox{}
\vskip 3.0in\relax\noindent\hskip 1.5in\relax
\includegraphics{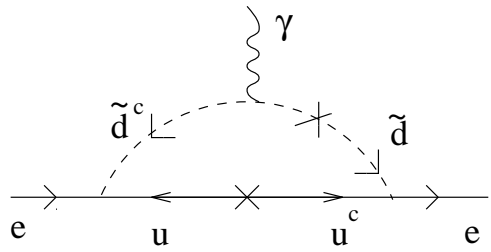} \vskip .25in
\caption{   One loop  diagram       contributing to   electric dipole moment 
   of electron.}
\end{figure}
 
\end{document}